\documentclass{pasj00}
\draft

\begin{document}
\SetRunningHead{M.ando et al.}{LBGs at $z\sim5$: Rest-frame UV Spectra II}
\Received{2007/01/29}
\Accepted{2007/05/07}

\title{Lyman Break Galaxies at $z\sim5$: Rest-frame UV Spectra II \thanks{\rm Based on data collected at Subaru Telescope,
which is operated by the National Astronomical Observatory of Japan.}}

\author{Masataka \textsc{Ando},
Kouji \textsc{Ohta}}
\affil{Department of Astronomy, Kyoto University,\\
Oiwake-cho, Kirashirakawa, Sakyo-ku, Kyoto, 606-8502}
\email{andoh@kusastro.kyoto-u.ac.jp, ohta@kusastro.kyoto-u.ac.jp}
\author{Ikuru \textsc{Iwata}}
\affil{Okayama Astrophysical Observatory, National Astronomical Observatory of Japan,\\
Kamogata, Okayama 719-0232}\email{iwata@oao.nao.ac.jp}
\author{Masayuki \textsc{Akiyama}, Kentaro \textsc {Aoki},
and Naoyuki \textsc{Tamura}}
\affil{Subaru Telescope, National Astronomical Observatory of Japan, \\
650 North A'ohoku Place, Hilo, Hawaii 96720, USA}
\email{akiyama@subaru.naoj.org, kaoki@subaru.naoj.org, naoyuki@subaru.naoj.org}

%

\KeyWords{galaxies: evolution --- galaxies: formation --- galaxies:
high-redshift}

\maketitle

\begin{abstract}
We present the results of spectroscopy of
Lyman Break Galaxies (LBGs) at $z\sim5$ in the J0053+1234 field
with the Faint Object Camera and Spectrograph on the Subaru telescope.
Among 5 bright candidates with $z' < 25.0$ mag, 2 objects are confirmed to
 be at $z\sim5$ from their Ly$\alpha$ emission and the continuum depression
 shortward of Ly$\alpha$.
The EWs of Ly$\alpha$ emission of the 2 LBGs are not so strong 
to be detected as
 Ly$\alpha$ emitters, and one of them shows strong low-ionized
 interstellar (LIS) metal absorption lines.
Two faint objects with $z' \geq 25.0$ mag are also
 confirmed to be at $z\sim5$, and their spectra show strong Ly$\alpha$
emission in contrast to the bright ones.
These results suggest
a deficiency of strong Ly$\alpha$ emission in bright LBGs at
 $z\sim5$, which has been discussed in our previous paper.
Combined with our previous spectra of LBGs at $z\sim5$
obtained around the Hubble Deep Field-North (HDF-N),
we made a composite spectrum of UV luminous 
($M_{1400} \leq -21.5$ mag) LBGs at $z\sim5$.
The resultant spectrum shows a weak Ly$\alpha$ emission and 
strong LIS absorptions which suggests that the bright LBGs at $z\sim5$ 
have chemically evolved at least to $\sim$0.1 solar metallicity.
For a part of our sample in the HDF-N region, we obtained near-to-mid infrared
 data, which constraint stellar masses of these objects.
With the stellar mass and the metallicity estimated
 from LIS absorptions, 
the metallicities of the LBGs at $z\sim5$ tend to be lower
 than those of the galaxies with the same stellar mass at $z \lesssim 2$,
 although the uncertainty is very large.
\end{abstract}

\section{Introduction}

Spectroscopic studies of high-redshift ($z>2\sim3$) 
galaxies have been advanced recently.
One of the largest spectroscopic samples of high-z galaxies is
obtained by follow-up spectroscopy for Lyman Break galaxies (LBGs: e.g.,
\cite{Ste03}) which are selected by using multi-band photometric data.
So far, about a thousand of optical spectra have been obtained each for
LBGs at $z\sim3$ (e.g., \cite{Shap03})
and star-forming galaxies at $z\sim2$ 
(BM/BX objects: e.g., \cite{Ste04}).
Intensive spectroscopic studies based on magnitude-limited samples 
have also identified several hundreds of galaxies at $z\sim2-4$
(e.g. FORS Deep Field (FDF) survey: \cite{Noll04}, VIRMOS-VLT Deep
Survey: \cite{Lefe05}).

These studies revealed  spectroscopic features of galaxies 
at $z\sim2-3$ in their rest-frame UV wavelength region, and
deepened our understanding of their nature.
\citet{Shap03} found relations between the
strength of the Ly$\alpha$ equivalent width (EW) and other spectroscopic
features (e.g., EW of the low-ionized interstellar (LIS) metal
absorption lines, UV continuum slope, velocity offset between
Ly$\alpha$ and LIS absorptions) for the LBGs at $z\sim3$,
which sheds light on properties of an interstellar matter 
such as a gas outflow, a velocity dispersion of the gas,
and a covering fraction of dust.
With a part of FDF data, \citet{Meh02} found that EW of the CIV 
$\lambda\lambda 1548, 1550$ absorption line decreases from $z\sim1.4$
to $\sim3.4$.
The decrease in the EW of CIV is confirmed with the full sample of FDF, 
and it seems to continue to $z\sim4$ \citep{Noll04}. 
Since the EW of CIV is an indicator of the metallicity 
in local starburst galaxies (e.g., \cite{Hek98}),
the result suggests a chemical evolution of galaxies in this
redshift range \citep{Meh02},
although the correlation between the metallicity and 
the EW of CIV has not been calibrated at high-redshift 
and thus is not ensured.
Near-infrared (NIR) spectra of galaxies at $z\sim2-3$ have also been 
obtained and give information about kinematics and 
metallicity of the galaxies.
The velocity offset between nebula emission lines and
interstellar absorption lines is generally seen in the LBGs, suggesting
the presence of the gas outflow (e.g., \cite{Pet01}).
From the emissions lines, gas metallicity in the galaxies can be derived 
by using metallicity estimators  (the $R_{23}$ and the N2 index);
for example, \citet{Tep00}, \citet{Kob00}, and \citet{Pet01} 
derived metallicity of LBGs at $z\sim3$ which ranges typically from
0.1$-$0.9 solar value.
\citet{Shap04} showed that majority of the seven NIR bright 
($K_{s,Vega}$ $\leq 20$ mag)
BX galaxies have near the solar metallicity, and
\citet{Erb06} investigated the extended sample
mainly with $K_{s,Vega}$ $< 21.5$ mag 
and found the median metallicity of 87 galaxies at $z\sim2$ to be about
0.5$-$0.6 solar value.

Recently, surveys of galaxies at $z\gtrsim5$ have been carried out
(e.g. 
\cite{Iwa03,Lehn03,Sta03,Dick04,Ouchi04a,Shima05,Yan05,Bou06a,Yoshida06}),
and challenges of finding galaxies even at
$z\sim7-10$ have also started 
(e.g., \cite{Pello04,Kneib04,Bou05,Bou06b,Mann07}).
However, the progress of optical follow-up spectroscopy for galaxies at
$z\gtrsim4-5$ is much slower than that for galaxies at $z=2\sim3$
because objects are getting fainter, and characteristic features of 
UV spectrum of
star-forming galaxies (e.g., Ly$\alpha$, LIS absorptions)
redshift to the wavelength region where night sky emissions
are very strong.
Although attempts of spectroscopic observations of galaxies at
$z\gtrsim5$ were made
(e.g., \cite{Spi98,Dey98,Wey98,Lehn03,Sta04}),
most of the galaxies were identified only with a Ly$\alpha$ emission
because of the low signal-to-noise ratio of their continuum light.

We also started an optical spectroscopy of LBGs at $z\sim5$.
Targets were selected from our surveys for LBGs at $z\sim5$ 
(\cite{Iwa03}, 2007) which consist of 
the deep and wide $V, I_C$, and $z'$-band imaging survey with 
the Subaru/Suprime-Cam in the two target fields: the region including 
the Hubble Deep Field-North (the HDF-N region) and the J0053+1234 region.
Our survey fields are quite wide (1290 arcmin$^2$ in total), 
and we obtain 228 bright objects with $z'_{AB}<25.0$ mag
($L\gtrsim L^{\ast}$ in UV luminosity function (UVLF)
of LBGs at $z\sim5$).
We reported an initial result of the spectroscopy for a part of 
the objects in the HDF-N region (\cite{Ando04}: hereafter Paper I).
Thanks to the object brightness and good observing conditions, 
we detected continuum features of the LBGs at $z\sim5$ 
and identified the galaxies to be at $z\sim5$
without Ly$\alpha$ emission line.
The most striking result was that there is no or a weak 
(EW$_{\rm rest}<$ 10\AA ) Ly$\alpha$ emission line in 7 bright LBGs
identified.
Inspired by the result, the deficiency of the strong Ly$\alpha$ 
emission in the bright LBGs at $z\sim5$
was investigated by using combined data of Paper I,
our new results obtained in the J0053+1234 region, 
and those from the literature;
the deficiency is generally seen in the LBGs at $z\sim5$
and even at $z\sim6$ \citep{Ando06}.
If the quenched Ly$\alpha$ emission is due to the dust, this deficiency
suggests that bright LBGs have larger amount of dust and higher
metallicity than faint ones.
This hypothesis suggests that bright LBGs experienced star formation
earlier than that of faint ones: differential evolution of the 
high-redshift galaxies with their luminosity, 
although there are other possible origins of the deficiency
(e.g., a velocity structure in/around the galaxy).

In this paper, we present the data of spectroscopic observation of
LBGs at $z\sim5$ in the J0053+1234 region which were used
in our previous paper \citep{Ando06}.
In Section 2, we describe a sample selection, observation, and
data reduction.
The features of obtained spectra of the LBGs at $z=4-5$ in the J0053+1234
region are shown in Section 3.
Combining this result with our previous results (Paper I),
we described spectroscopic properties of bright LBGs at $z\sim5$ and 
metallicity against stellar mass in the section.
Throughout this paper, we adopt flat $\Lambda$ cosmology, $\Omega_M=0.3$,
$\Omega_{\Lambda}=0.7$, and $H_0=70$ km s$^{-1}$Mpc$^{-1}$. 
The magnitude system is based on AB magnitude \citep{Oke83}.

\section{Observation and Data Reduction}
The photometric sample of LBGs at $z\sim5$ in the J0053+1234 region
was selected based on a deep and wide broad-band ($V, I_C,$ and $z'$)
imaging survey by using Subaru/Suprime-Cam \citep{Iye04,Miya}:
a total of 114 (236) candidates with $z' <$ 25.0 (25.5) mag in an effective
survey area ($\sim$800 arcmin$^2$). 
Details of imaging observations and the color selection are described in
\citet{Iwa07}.
In order to detect continuum and absorption feature,
we selected the bright  ($z' < 25.0$) LBG candidates at $z\sim5$
as the main spectroscopic targets.
Because the entire survey field was too wide to obtain spectra for the whole
sample of these LBG candidates with a limited field of view ($6'\phi$)
of the Faint Object Camera and Spectrograph 
(FOCAS: \cite{kashi})
on the Subaru telescope and limited observing time, 
we selected a target MOS field to contain the bright LBGs as many as possible.
When a slit of a bright target overlapped with a slit of another bright one,
we chose the object with a higher central surface brightness.
Finally, we designed one FOCAS mask which covers 5 bright LBGs.
The mask also covered 8 fainter ($z' \geq 25.0$ mag) LBGs
which are in the gaps of the slits of main targets on the mask.

Spectroscopic observations with the Subaru/FOCAS were made on 2004 
September 14
under a clear condition, and a seeing during the observing run 
was typically $\sim 0.^{\prime\prime}$8 (FWHM).
We used a grism of 300 lines mm$^{-1}$ blazed at 7500 \AA\ 
and the SO58 order-cut filter (the same setting as Paper I),
which gave wavelength coverage from 5800\AA\ to 10000\AA\
(depending on a slit position on the mask) with a pixel scale of 1.34\AA.
The slit lengths were typically 10$^{''}$, and the slit widths were
fixed to be 0.$^{''}$8, giving a spectral resolution of R$\sim$700
which was measured with night sky emission lines.
An exposure time of each frame was 0.5 hours, and a total effective
exposure time was 6 hours.

The data were reduced by using IRAF\footnote{Image
Reduction and Analysis Facility, distributed by National Optical
Astronomical Observatories, which are operated by the Association of
Universities for Research in Astronomy, Inc., under cooperative
agreement with the National Science Foundation.}
with the same manner as Paper I except for the re-subtraction of the
night sky emissions.
We subtracted bias from object frames by using 
an overscan region of each frame
and averaged bias frame in order to remove the characteristic 
bias-pattern of FOCAS CCDs.
Flat-fielding was made with normalized average of dome-flat images. 
The wavelength calibration was made by using night-sky emission lines.
The RMS error of the wavelength calibration was $\sim0.7$\AA.
After the sky subtraction, 
the spectrum of each object was carefully aligned and averaged. 
Because some objects show too low continuum S/N to subtract sky emission
in each frame,
we carried out sky subtraction after combining spectra for these objects.
Five pixels were binned toward wavelength direction in order to improve
S/N, and one-dimensional spectra were obtained
with APALL task of IRAF.
Since our objects were faint, we did not fix the aperture width
for the extraction
but we determined it by eye for each object to trace it well.
The flux calibration and the sensitivity correction were made for the final
spectra by using standard stars of Wolf1346 and PG0823+546.
The correction for the atmospheric A and B bands was also made by tracing
those absorption features in the spectra of standard stars.
For some of the objects, the sky subtraction was not well achieved
due to the low-S/N of the continuum or the presence of the bad column of
the CCDs.
Thus we re-subtracted the scaled night sky emissions from the
one-dimensional spectrum of such objects.

\section{Results and Discussions}

\subsection{Redshift Determination}

Among 5 bright targets, we identified 2 objects to be LBGs at $z=4-5$,
and the resultant spectra are shown in Figure 1 (No.1: top left, No.2:
top right).
LBG identifications were mainly made by the presence of 
the continuum depression shortward of the redshifted
Ly$\alpha$ due to the inter-galactic HI absorption
and the Ly$\alpha$ emission line near the Ly$\alpha$ depression.
In addition, the Ly$\alpha$ emission line of each object shows an asymmetric
shape which is considered to be characteristic feature of LBGs and Ly$\alpha$
emitters (LAEs) at high-redshift.
Note that emission-like features in the spectrum of object No.1 at
$\sim$6300\AA\ and in the wavelength region longer than 7300\AA\ 
are the residuals of the night sky emissions since the sky subtraction was
not well achieved due to the bad column in the CCD.
For object No.2, we identified some LIS absorption lines
(SiII $\lambda$1260, OI+SiII $\lambda$1303, and CII $\lambda$1335) 
at almost the same redshift of the Ly$\alpha$
which are characteristic features of nearby star-forming galaxies
(e.g., \cite{Hek98}) and LBGs at $z\sim3-5$
(e.g., \cite{Ste96a,Shap03,frye}; Paper I; \cite{Iwa05}),
supporting the identifications of high redshift star-forming galaxies.
Redshifts of these two bright LBGs were determined from the peak of
Ly$\alpha$ emission:
4.797 and 4.267 for object No.1 and No.2, respectively.
The error of the redshift determination was $\sim0.005$.
The rest of the bright sample were not identified to be at $z\sim5$
because of their low S/N of the spectra.
These objects tend to have lower central surface 
brightness compared with the confirmed objects.

In addition to the bright targets, 
we identified 2 objects among faint bonus targets with
$z'\geq25.0$ mag to be LBGs at $z=4-5$. 
Spectra of them are shown in Figure 1 (No.3: bottom
left, and No.4: bottom right).
These objects show a strong (EW$_{\rm obs}\sim220$\AA\ and $\sim$420\AA\ 
for object No.3 and No.4, respectively) single emission line;
peak of the emission corresponds to the redshift of 4.391 and 4.491
for object No.3 and No.4, respectively if this emission is Ly$\alpha$.
A clear asymmetry of the emission line profile
and possible continuum depression shortward of the emission
are seen in the spectrum of object No.4, 
which supports this object is really at $z=4-5$.
Because the redshift determinations for these objects were based only on
the single emission line, we examine the possibility that they are
foreground emission galaxies.
For example, if object No.3 (No.4) is an [OII] $\lambda3727$
emitter at $z=0.76$ (0.79),
they should show the redshifted H$\beta$ emission line at $\sim$8550(8710)\AA\
and the [OIII]$\lambda$5007 emission line 
at $\sim$8810(8970)\AA\ with EW$_{\rm obs}$ of $\sim$70(130)\AA\ and
EW$_{\rm obs}$ of $\sim$110(210)\AA, respectively by assuming the median of 
the observed line ratios of star-forming galaxies at $z=0.3\sim1.0$
(0.3 and 0.5, respectively: \cite{Kob04}).
As shown in Figure 1, emission-like
features are seen around at 
a part of the expected wavelengths in both spectra. 
However, these features are residual night-sky emissions, and 
there is no other significant
(EW$_{\rm obs}\gtrsim100$\AA) emission lines at the expected wavelengths.
Thus these strong emissions are considered to be the Ly$\alpha$ emission
from LBGs at the target redshift.
We can not completely rule out the possibility that they are 
foreground emission galaxies by only this analysis,
because some of the expected lines of foregrounds are in the wavelength
region of severe night sky emissions, and the line ratios have wide
object-to-object variances \citep{Kob04}.
But the $V-I_C$ colors of these objects (1.98 mag and 1.85 mag) 
are too red to regard them as star-forming galaxies at $z<1$; 
we calculated the SEDs of star-forming galaxies at $z<1$ by using 
the population synthesis code P\'{E}GASE version 2 \citep{pegase} 
with Salpeter IMF \citep{Sal55} and by
assuming constant star formation with an age of 100 Myr,
which gives their $V-I_C$ colors in observed-frame
of $-0.4-0.1$ mag and $0.2-0.8$ mag for $E(B-V)=0.0$ mag and
$E(B-V)=0.4$ mag, respectively with the 
dust extinction curve by \citet{Cal00}.
Thus we consider both of the objects are at $z=4-5$.

\begin{figure}
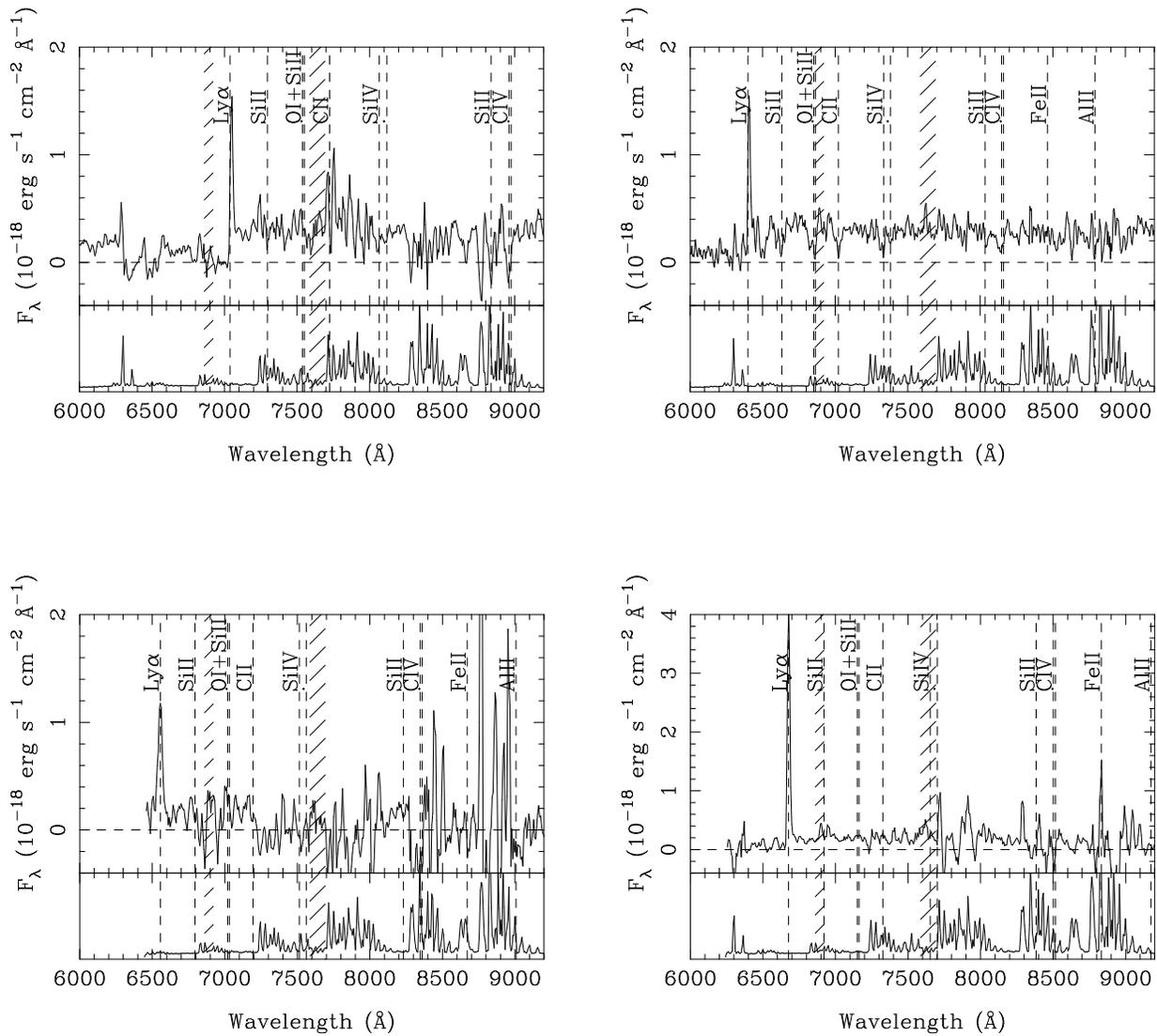

\begin{center}
\rotatebox{270}{\resizebox{6.5cm}{!}{\FigureFile(70mm,50mm){f1a.ps}}}
\hspace{0.5cm}
\rotatebox{270}{\resizebox{6.5cm}{!}{\FigureFile(70mm,50mm){f1b.ps}}}\\
\vspace{1.5cm}
\rotatebox{270}{\resizebox{6.5cm}{!}{\FigureFile(70mm,50mm){f1c.ps}}}
\hspace{0.5cm}
\rotatebox{270}{\resizebox{6.5cm}{!}{\FigureFile(70mm,50mm){f1d.ps}}}\\
\caption{Spectra of LBGs at $z=4-5$ in the J0053+1234 region. 
 {\it Top}: Object No.1 ({\it left}) and No.2 ({\it
 right}). {\it Bottom}: Object No.3 ({\it left}) and No.4 ({\it right}).
These spectra are smoothed with a boxcar over 3-pixel. 
Positions of line
 features seen in LBGs at $z=3$ and local starburst galaxies 
(e.g., \cite{Shap03,Hek98}) are
 shown with vertical dashed lines.
A scaled sky spectrum is shown in a lower part of each panel, and the
 atmospheric A-band and B-band absorptions are shown as vertical hatched
 regions.}
\end{center}
\end{figure}

\subsection{Color Selection and Redshift Distribution of LBGs at $z\sim5$}

Figure 2 shows the positions of observed objects in the $V - I_C$
and $I_C - z'$ two-color diagram from the present results and Paper I.
Photometric properties of spectroscopically identified LBGs at $z=4-5$
in the J0053+1234 region are summarized in Table 1.
Filled circles show LBGs at $z=4-5$ in this study: 2 bright objects
as large circles and 2 faint objects as small ones.
Filled squares show bright LBGs at $z\sim5$ in our previous results
(Paper I).
Open triangles and open-inverse triangles show bright LBG candidates 
unidentified in our present observation and the previous results, 
respectively.
So far, no bright spectroscopic targets (total of 22 LBG candidates)
have been identified as foreground galaxies, though a half of them
were not identified due to the low-S/N of their spectra.
In order to examine our color selection criteria further, we also show 
the positions of the Galactic M stars which were identified in our
previous observations as filled pentagons in Figure 2.
Our selection 
criteria for LBGs at $z\sim5$ are reasonable to exclude interlopers,
though more large and systematic spectroscopic observations are desirable.

\begin{table}
\begin{center}
\caption{Photometric properties of the LBGs identified to be at $z\sim5$ in the J0053+1234 region.\label{tbl-1}}
\begin{tabular}{lccccccc}
\hline\hline
No. (ID) & RA(J2000) & DEC(J2000) &
$z'$\footnotemark[$*$] & $V-I_C$\footnotemark[$\dagger$] &
$I_C-z'$\footnotemark[$\dagger$] \\
\hline
1 (106426) & 00:52:21.34 & +12:32:35.3 & 24.07 & 2.59 & 0.23 \\
2 (104115) & 00:52:43.27 & +12:32:08.2 & 24.22 & 1.75 & 0.16 \\
3 (091813) & 00:52:39.88 & +12:29:44.1 & 25.03 & 1.98 & 0.06 \\
4 (093014) & 00:52:37.37 & +12:29:58.7 & 25.26 & $>$1.85 & 0.16 \\
\hline
\multicolumn{6}{@{}l@{}}{\hbox to 0pt{\parbox{160mm}{\footnotesize
\par\noindent
\footnotemark[$*$] $z'$ magnitude. MAG\_AUTO from SExtractor \citep{Bet}
 is adopted. \\ 
\footnotemark[$\dagger$] These values are measured with a
 $1^{\prime\prime}.6$ aperture. \\
}\hss}}
\end{tabular}
\end{center}
\end{table}

\begin{figure}  
\begin{center}
\FigureFile(160mm,100mm){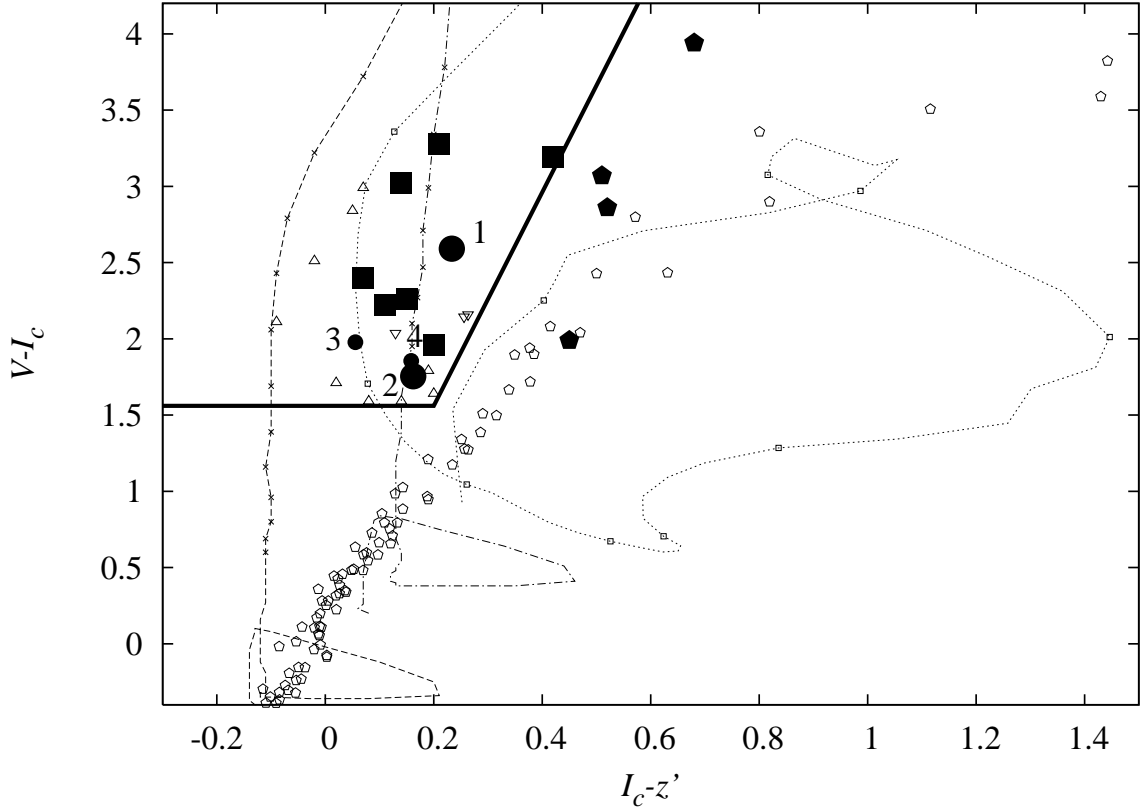}
\end{center}
\caption{
Positions of identified objects in the $I_C-z'$ and $V-I_C$ two-color
 diagram. 
Our color selection criteria for LBGs at $z\sim5$ (\cite{Iwa03}; 2007)
are indicated by thick lines.
The objects confirmed to be at $z\sim5$ in this study are shown as
filled circles with each identification number in Table 1.
The large circles show 2 bright objects, and small ones show 2 faint
 objects.
Filled squares represent 7 bright LBGs at $z\sim5$ from Paper I. 
Open triangles and open-inverse triangles show bright LBG
candidates unidentified in our present observation and the previous
 results (Paper I), respectively.
Filled pentagons show objects identified to be foreground
 objects (Galactic M stars) from Paper I.
A dashed (dot-dashed) line represents a color track of a model LBG
 spectrum with $E(B-V)=0.0$ mag ($E(B-V)=0.4$ mag) from \citet{Iwa07}.
Small crosses are plotted on the tracks for $z \geq 4.0$ with an
 interval of 0.1.
 Model SEDs of LBGs are generated by using the population synthesis code
 P\'{E}GASE version 2 \citep{pegase} with Salpeter IMF \citep{Sal55} and by
 assuming constant star formation with an age of 100 Myr.
IGM absorptions at each redshift are calculated by using the analytic
formula by \citet{Inoue05}.
Calzetti's dust extinction curve \citep{Cal00} is assumed 
to make model spectra with $E(B-V)=0.4$ mag.
A dotted line refers to a color track of an elliptical galaxy \citep{cww}. 
Small open squares are plotted on the track with a redshift interval of 0.5. 
Small open pentagons indicate the colors of A0 -- M9
 stars calculated based on the library by \citet{Pick}.
} 
\end{figure} 

The redshift distribution of the spectroscopically confirmed LBGs at $z\sim5$
in this study and Paper I is shown in Figure 3.
In Figure 3, we plot the expected redshift distribution of our 
sample of LBGs at $z\sim5$ as a solid line normalized at $z=4.7$.
The distribution is calculated based on the expected detection rates
of the LBGs against the apparent magnitude and the redshift for each
survey field (figure 6 by \citet{Iwa07}).
However, the detection rates were derived through simulations
in the two-color diagram to estimate the completeness of the photometric
catalog, and they do not include the difference 
of survey volumes for each redshift bin
and the difference of the number density at the corresponding UV
magnitude for an apparent magnitude at each redshift bin.
In order to derive the expected distribution and compare it with
the observed redshift distribution of the spectroscopic sample,
we corrected the detection rates
by adopting the redshift bin size of $\Delta z=0.2$
and assuming the UVLF of the $z\sim5$ LBGs \citep{Iwa07}.
Since almost all our spectroscopic sample covers $z'$ magnitude brighter
than 25.5 mag, we calculated the expected redshift distributions for
three magnitude ranges of 24.0$-$24.5 mag, 24.5$-$25.0 mag, and
25.0$-$25.5 mag for each field, and took the average of them.
The distribution of spectroscopically confirmed LBGs is broadly
consistent with the expected distribution, but 
there seems to be a deficiency of the LBGs at around
$z\sim5.0$.
Although the sample size is still small and the deficiency may
not be significant,
this may be due to the presence of severe night sky emission lines
which prevent us from detecting the continuum depression 
and/or Ly$\alpha$ emission.
The night sky emission lines whose 
wavelengths are converted to the redshifts corresponding to 
the redshift of Ly$\alpha$ are also shown with dotted line
in the figure; the sky emission lines are severe for the
redshifts larger than 4.9.

As described in section 3.1, spectroscopically identified LBGs in 
the J0053 region have redshifts of $4.3-4.8$ with a mean of 4.5, 
which is slightly lower than that of the objects in Paper I 
with redshifts of $4.5-5.2$ (mean value of 4.7).
In fact, as seen in Figure 2, 
the 4 objects in the J0053 region are bluer in 
the $V-I_C$ color than the 7 objects in Paper I.
The mean $V-I_C$ color of spectroscopic sample is 2.62 mag and 2.04 mag 
for the objects in the HDF-N region and in the J0053 region, respectively.
However the difference may be due to the variation among small
numbers of objects, and we should have more objects with spectroscopic
redshifts to examine this point further.

\begin{figure}
\begin{center}
\FigureFile(140mm,80mm){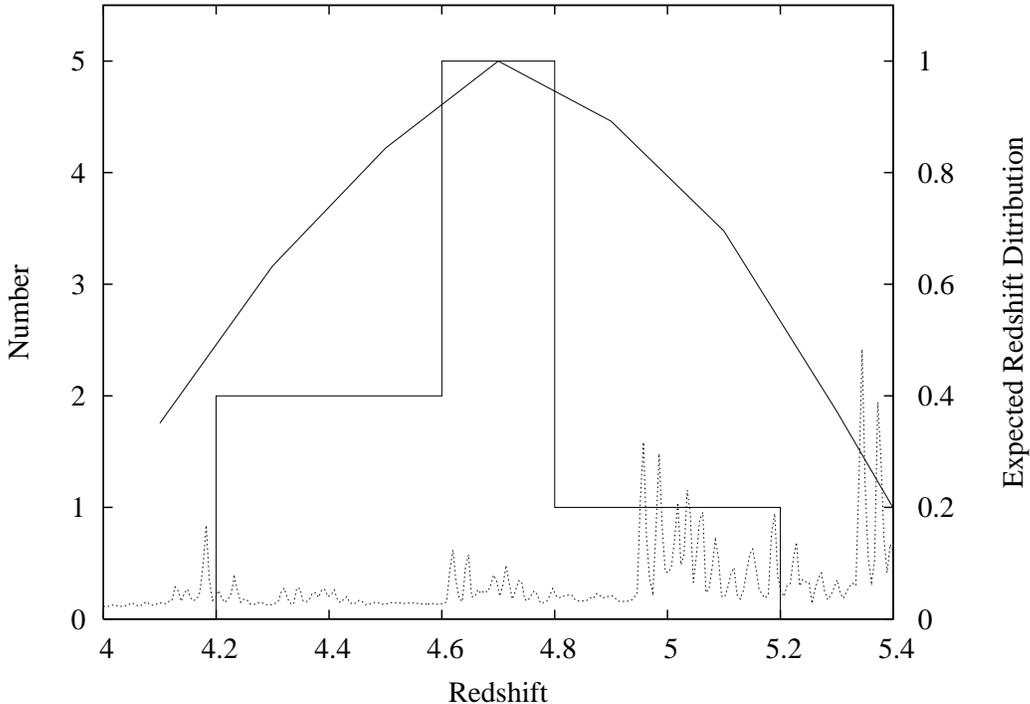}
\end{center}
\caption{Redshift distribution of LBGs at $z\sim5$ from this study and
 our previous spectroscopic results (Paper I).
The night sky emission lines whose wavelengths are converted to the
redshifts corresponding to the redshift of Ly$\alpha$
are also shown with dotted line.
Expected redshift distribution normalized at $z=4.7$ for LBGs at $z\sim5$
is also plotted as solid line (right vertical axis,
see text for the details).
} 
\end{figure} 

\subsection{Spectra of LBGs at $z\sim5$ in the J0053+1234 region}

As shown in Figure 1, spectra of bright LBGs in the J0053+1234
region show Ly$\alpha$ and LIS absorptions (SiII $\lambda$1260, OI+SiII
$\lambda$1303, and CII $\lambda$1335) which are typical line features of
star-forming galaxies at low and high redshifts.
The Ly$\alpha$ EW$_{\rm rest}$s of object No.1 and No.2 are 14$\pm3$\AA\
and 21$\pm4$\AA , respectively, and their observed fluxes are
2.5$\times 10^{-17}$ erg s$^{-1}$ cm$^{-2}$ and 2.4$\times 10^{-17}$ 
erg s$^{-1}$ cm$^{-2}$, respectively
corresponding to the luminosities of 5.8$\times 10^{42}$ erg s$^{-1}$ and 
4.3$\times10^{42}$ erg s$^{-1}$, respectively.
These values of Ly$\alpha$ do not include absorption component and are not
corrected for the IGM absorption. 
The errors of EW values in this paper include
the uncertainty in determination of the continuum level
and the RMS error of the continuum.
The values of EW of Ly$\alpha$ are larger than the average value of 
our previous results for 7 bright LBGs (2.5$\pm$3.7\AA , Paper I), 
but they are not so large to be detected as definitive LAEs.
Typical Ly$\alpha$ EW$_{\rm rest}$ value for selecting LAEs is 
15$\sim$20\AA .
In contrast to the result for the bright targets,
the Ly$\alpha$ emissions of the faint LBGs are strong; 
EW$_{\rm rest}$ of object No.3 and No.4 are 40$\pm12$\AA\ and 
77$\pm23$\AA , respectively.
The observed fluxes of Ly$\alpha$ are 
3.7$\times 10^{-17}$ erg s$^{-1}$ cm$^{-2}$ and 8.3$\times 10^{-17}$
erg s$^{-1}$ cm$^{-2}$, respectively corresponding to the luminosities of
7.1$\times 10^{42}$ erg s$^{-1}$ and 1.7$\times10^{43}$ erg s$^{-1}$,
respectively.
We summarize the observed quantities in Table 2.

The bright LBGs at $z\sim5$ tend to show no or weak Ly$\alpha$ emission;
average EW$_{\rm rest}$ of Ly$\alpha$ emission of the total of 9 bright
LBGs is $\sim5.8\pm7.2$\AA , while
strong Ly$\alpha$ emission is seen among the faint LBGs.
Because we can not deny the presence of faint LBGs with no or weak
Ly$\alpha$ emission from the current data, these results suggest, 
at least, a deficiency of strong Ly$\alpha$ emission in bright LBGs at
$z\sim5$ and might suggest the 
UV luminosity dependence of the Ly$\alpha$ emission.
Combining with other spectroscopic results from the literature,
we found that the deficiency is generally seen in LBGs at $z\sim5$
and possibly in LBGs at $z\sim6$ \citep{Ando06}.
The trend is also seen in the galaxies at $z=4\sim6$ 
(Vanzella et al. 2006; 2007).
Possible origins of the deficiency of strong Ly$\alpha$ emission 
are discussed by \citet{Ando06}.
The deficiency or the UV luminosity dependence of the Ly$\alpha$
are also seen in the recent results of spectroscopy of LBGs at $z\sim3$;
\citet{Shap06} showed that 
UV luminous ($M_{UV}\sim-21.6$ mag) LBGs at $z\sim3$ tend to show
weaker Ly$\alpha$ emission than the total LBG sample of \citet{Shap03}.
From the re-binned sample of \citet{Shap03} with their UV luminosity, 
\citet{Keel06} showed that Ly$\alpha$ emission is
weaker in the UV luminous ($M_{UV}< -21.0$ mag) LBGs than in fainter LBGs.

For object No.2, we identified LIS absorption lines
(SiII $\lambda$1260, OI+SiII $\lambda$1303, and CII $\lambda$1335) 
at almost the same redshift as the Ly$\alpha$ emission.
We also identified SiIV $\lambda$1394 (EW$_{\rm rest}=-2.0\pm1.1$ \AA )
\footnote{We resolved SiIV $\lambda\lambda$1394,1403 doublet, 
but $\lambda$1403
component was not detected significantly (comparable to 1$\sigma$),
which is consistent with the observed ratio for LBGs at $z\sim3$ 
\citep{Shap03}.}
and CIV $\lambda\lambda$1548,1551 (EW$_{\rm rest}=-4.9\pm1.4$\AA )
which are also seen in the spectra of LBGs at $\sim3$ 
(e.g., \cite{Shap03}).
The average EW of the three LIS absorption lines is $-4.2\pm1.4$\AA\
which is larger than the average value of our previous result 
($-2.8\pm1.2$\AA\ : Paper I).
The peak of Ly$\alpha$ emission line is redshifted by $280 \pm 210$ 
km s$^{-1}$ relative to the average of the LIS lines.
The peak-to-valley velocity offset which may be related to 
a large scale gas outflow was also reported in LBGs at
$z=3\sim5$ (e.g., \cite{Shap03, frye}; Paper I).
The offset value of object No.2 is smaller than typical value 
of LBGs at $z\sim3$ ($500-600$ km s$^{-1}$: \cite{Shap03}) 
and that of LBGs at $z\sim5$ ($\sim600$ km s$^{-1}$: Paper I).
For object No.1, we could not identify significant absorption
features due to the residuals of the sky subtraction.

\begin{table}
\begin{center}
\caption{Spectroscopic properties of the LBGs identified to be at $z=4-5$ in the J0053+1234 region.\label{tbl-2}}
\begin{tabular}{lccccccc}
\hline\hline
No. (ID) & Redshift\footnotemark[$*$] &
EW(Ly$\alpha$)\footnotemark[$\dagger$] & 
$F_{\rm Ly\alpha}$\footnotemark[$\ddagger$] &
$L_{\rm Ly\alpha}$\footnotemark[$\S$] \\
\hline
1 (106426) & 4.797 & 14$\pm3$ & 2.5$\times 10^{-17}$  & 5.8$\times 10^{42}$ \\
2 (104115) & 4.267 & 21$\pm4$ & 2.4$\times 10^{-17}$ & 4.3$\times 10^{42}$ \\
3 (091813) & 4.391 & 40$\pm12$ & 3.7$\times 10^{-17}$ & 7.1$\times 10^{42}$ \\
4 (093014) & 4.491 & 77$\pm23$ & 8.3$\times 10^{-17}$ & 1.7$\times 10^{43}$ \\
\hline
\multicolumn{5}{@{}l@{}}{\hbox to 0pt{\parbox{110mm}{\footnotesize
\par\noindent
\footnotemark[$*$] Redshifts are determined from the peak of
 Ly$\alpha$ emission. 
 The error of the redshift determination is $\sim0.005$. \\
\footnotemark[$\dagger$] Rest-frame equivalent width of Ly$\alpha$ emission in
 units of angstroms, not including Ly$\alpha$ absorption. 
EWs are taken to be positive for emission lines.
The error is estimated from
the uncertainty of the continuum level determination and the RMS error of
the continuum. \\
\footnotemark[$\ddagger$] Observed flux of the Ly$\alpha$ emission line in
 units of erg s$^{-1}$ cm$^{-2}$. \\
\footnotemark[$\S$] Luminosity of the Ly$\alpha$ emission line 
in units of erg s$^{-1}$.
The slit correction is not applied to the values. \\
}\hss}}
\end{tabular}
\end{center}
\end{table}

\subsection{Composite Spectrum of Luminous LBGs at $z\sim5$}

In order to overcome the low-S/N of each spectrum and investigate
typical spectroscopic properties of luminous LBGs at $z\sim5$, we produced a
composite spectrum of 8 LBGs at $z\sim5$ with $M_{1400}<-21.5$ mag
by averaging the spectra obtained in this study and our previous study.
We derived the absolute UV magnitude from the observed broad band
magnitude to the rest-frame 1400\AA\ magnitude assuming a continuum
slope $\beta$ ($f_{\lambda}\propto\lambda^{\beta}$) of $-1$ 
which is a typical value for LBGs at $z\sim3$ \citep{Shap03}.
Before co-adding the spectra, all wavelengths were converted to the
rest-frame scale using individual redshifts, and each flux density 
was normalized at 1250\AA .

Figure 4 shows the resultant composite spectrum re-binned to 
common dispersion of 1\AA\ pixel$^{-1}$.
Note that the emission-like features except for the Ly$\alpha$
(e.g., around 1275\AA , 1425\AA ) 
come from the residuals of sky emission lines, 
and these features are not real.
The continuum depression shortward of Ly$\alpha$ by IGM is clearly seen in
the composite spectrum.
A depression factor $D_A$ \citep{Oke82} for the composite spectrum 
is $\sim$0.6, which agrees with the value of 
the composite spectrum of LBGs at $z\sim5$ (Paper I) and the values of
QSOs at $z\sim5$ ($0.1-0.7$: e.g., \cite{Son02,Son04}).
As expected from the previous subsection, EW$_{\rm rest}$ of the
Ly$\alpha$ emission is not large (7.6$\pm0.8$\AA ), and 
LIS absorptions are also seen clearly.
The EWs of LIS absorptions are $-2.7\pm0.36$\AA , $-1.8\pm0.30$\AA , 
and $-1.5\pm0.29$\AA\
for SiII $\lambda$1260, OI+SiII $\lambda$1303, and CII $\lambda$1335,
respectively,
which suggests that bright LBGs have been chemically polluted to
some extent by metal at $z\sim5$.
The SiIV $\lambda\lambda$1394,1403 is also seen in the composite
spectrum, and the EW$_{\rm rest}$s of 
the SiIV $\lambda1394$ and the SiIV $\lambda1403$
are $-2.0\pm0.32$ \AA\ and $-1.0\pm0.26$ \AA , respectively.

The EW${_{\rm rest}}$ of Ly$\alpha$ is smaller than that of the composite
spectrum of LBGs at $z\sim3$ (15.1\AA : \cite{Shap03}); 
the EW values of LIS absorptions are comparable to or somewhat larger
than those of $z\sim3$
($-1.7$\AA , $-2.3$\AA , and $-1.5$\AA\
for SiII $\lambda$1260, OI+SiII $\lambda$1303, and CII $\lambda$1335,
respectively).
However, the sample luminosity range of our LBGs at $z\sim5$ 
($z' \lesssim 24.8$ mag corresponding to $M_{1400} \lesssim -21.5$ mag)
is different from that of LBGs at $z\sim3$
(mainly $R=23.5-25.5$ mag corresponding to $M_{1400}=-21.8\sim-19.8$ mag).
As described in the previous subsection, we found the deficiency of the
strong Ly$\alpha$ emission in luminous LBGs at $z\sim5$ and pointed out
the possible presence of the UV luminosity dependence on the spectroscopic
features.
Thus we can not discuss the evolution of the spectroscopic feature of LBGs
from a simple comparison of the composite spectra.

\begin{figure}
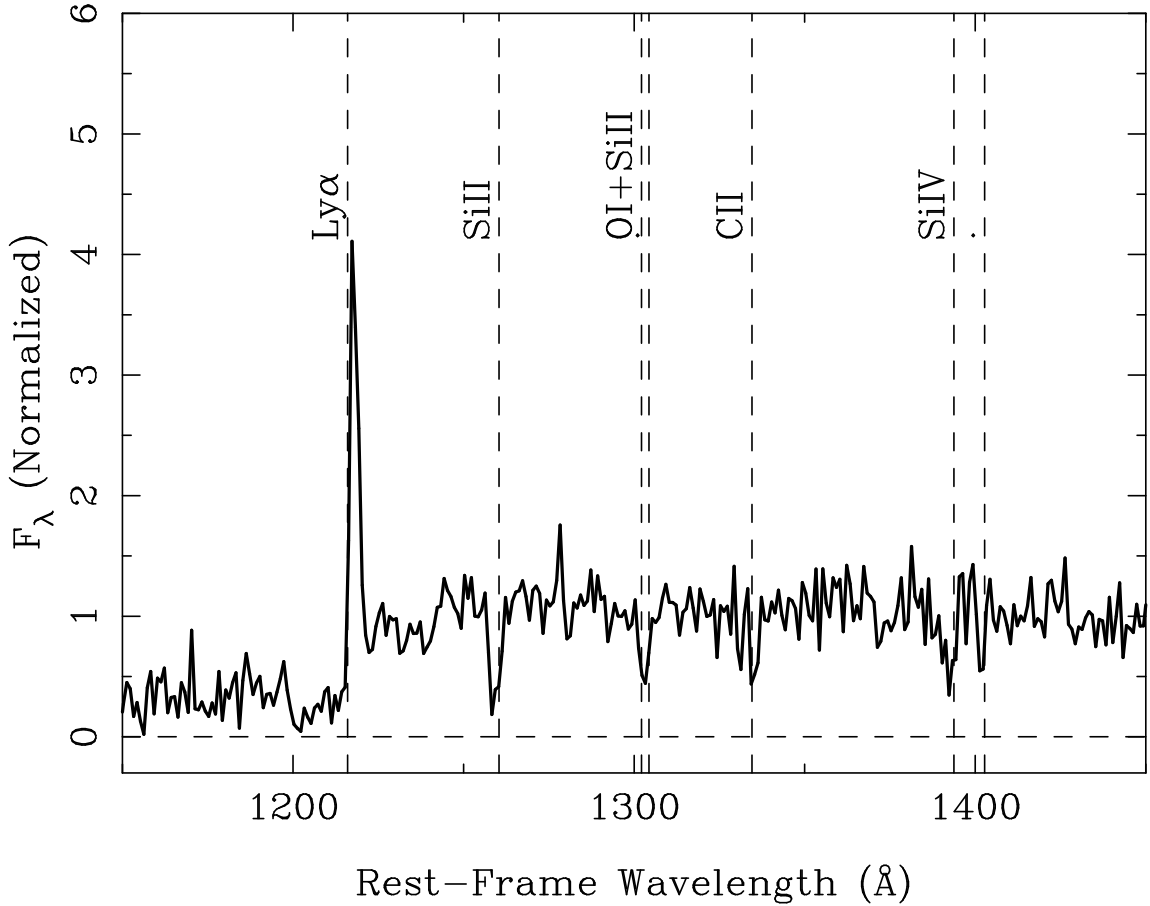
  
\begin{center}
\rotatebox{270}{\resizebox{12cm}{!}{\FigureFile(140mm,80mm){f4.ps}}}
\end{center}
\caption{
Composite spectrum of 8 luminous ($M_{1400}<-21.5$ mag) LBGs at
 $z\sim5$ from this study and Paper I. 
Note that most of emission-like features (e.g. around 1275\AA , 1425\AA ) 
come from the residuals of sky emission lines.
} 
\end{figure} 

Although the low S/N around the wavelength range
longer than 1400\AA\ does not allow us to estimate metallicity 
by using 1425
index (e.g., \cite{Rix}) and CIV index (e.g., \cite{Meh02}),
we roughly estimate mean metallicity of the 8 luminous LBGs at $z\sim5$ 
from LIS absorption lines of the composite spectrum assuming the
relation between the mean EW of the three LIS absorptions 
(SiII $\lambda$1260, OI+SiII $\lambda$1303, and CII $\lambda$1335)
and a metallicity calibrated with local star-forming galaxies \citep{Hek98}. 
Since the sample by \citet{Hek98} does not cover the low metallicity
(i.e., weak LIS absorptions) region, we extrapolated the EW(LIS)-metallicity
relation in the middle-to-high (12+log(O/H)$\gtrsim$8.0)
metallicity region to the lower metallicity region.
The estimated metallicity of 12+log(O/H) is about 7.6$\pm0.6$ 
corresponding to $\sim$0.1 (0.3$-$0.02) solar value (8.66: \cite{Asp04}).
The errors come from the uncertainty of the EW measurement as well as
that of the empirical relation between the EW(LIS) and metallicity.
It should be worth noting that the LIS lines are saturated, 
and EW(LIS) reflects not only the metallicity but also the velocity 
structure and covering fraction of the gas (e.g., \cite{Hek98,Shap03,Noll04}).
\citet{Erb06} show a composite rest-frame UV spectrum 
of objects at $z\sim2$ with a mean stellar mass of $7\times10^{10}M_{\odot}$
and mean metallicity of 12+log(O/H)$\sim8.55$
derived from N2 index method.
The EW(LIS) in the spectrum is $\sim2-3$\AA\
corresponding to 12+log(O/H) of $\sim7.6-8.2$ 
if we use  the local relation by \citet{Hek98}.
The result suggests that the local EW(LIS)-metallicity relation 
may systematically underestimate the metallicity of galaxies 
at high redshift.

\subsection{A Hint for Chemical Evolution of massive LBGs at $z\sim5$?}

We derived the individual metallicity for 8 LBGs at $z\sim5$ with the 
LIS absorption in this study and Paper I with the same manner mentioned above.
For the 3 objects in the HDF-N region among the 8 LBGs, 
Subaru/CISCO $K'$ band data and Spitzer/IRAC mid-IR data are available, 
and we can estimate their stellar masses by 
fitting their rest-frame UV to optical SEDs to the model SEDs.
(Details of NIR and MIR data of our LBGs at $z\sim5$
will be described elsewhere.)
The metallicities (12+log(O/H)) of
104268 (No.1 of Paper I), 144200 (No.2 of Paper I), 
and 148198 (No.8 of Paper I)
are 7.7$\pm 0.8$, 7.7$\pm 0.8$, and 8.7$\pm 0.9$, respectively.
For the stellar mass estimation, we adopted the same manner used in
\citet{Iwa05}.
We used the P\'{E}GASE version 2 population synthesis model \citep{pegase}
with Salpeter IMF \citep{Sal55}, constant star formation history, 
an the Calzetti's dust extinction curve \citep{Cal00} to make model spectra.
IGM absorptions were calculated for the given redshift
by using the analytic formula by \citet{Inoue05}.
Then the observed SEDs were fit to model SEDs, and the best fit model was
obtained by minimizing $\chi^2$. 
Resultant stellar masses (best fit value with 68\% confidence level)
of the objects 104268, 144200, and 148198 are 
5.82$^{+6.95}_{-3.93}\times10^{10}M_{\odot}$,
3.33$^{+0.94}_{-2.92}\times10^{10}M_{\odot}$, and
1.33$^{+0.67}_{-0.90}\times10^{11}M_{\odot}$, respectively. 

In Figure 5, open circles represent the stellar masses and the
metallicities of the three luminous LBGs at $z\sim5$, and
filled circle represents the average value of them.
Pluses, diamonds, and triangles show the data for galaxies at
$z\sim0.1$, $z\sim0.7$ and $z\sim2$, respectively \citep{Tre,Sav,Erb06},
which show the presence of mass-metallicity relation at each
redshift and its evolution from $z\sim2$ to $z\sim0$
\footnote{The metallicity of galaxies at $z\sim2$ \citep{Erb06} was
derived from N2 index (e.g., \cite{Pet04}) which is different from
methods used by \citet{Tre} and \citet{Sav} and may have a problems in
calibration (e.g., \cite{Erb06})
which may partly be a cause for the offset.}.
Although the uncertainties are quite large (and may contain some
systematics),
and our way of metallicity
estimation is different from those for galaxies at other redshifts, 
the metallicities of the LBGs at $z\sim5$ tend to be lower than those of the
galaxies with the same stellar mass at $z\lesssim2$
\footnote{The stellar masses by \citet{Erb06} were derived by using 
Chabrier IMF.
Thus our data points should be $\sim$1.8 times smaller (e.g., \cite{Erb06})
in Figure 5 in order to compare them with the data by \citet{Erb06}.}.

\begin{figure}  
\begin{center}
\FigureFile(140mm,80mm){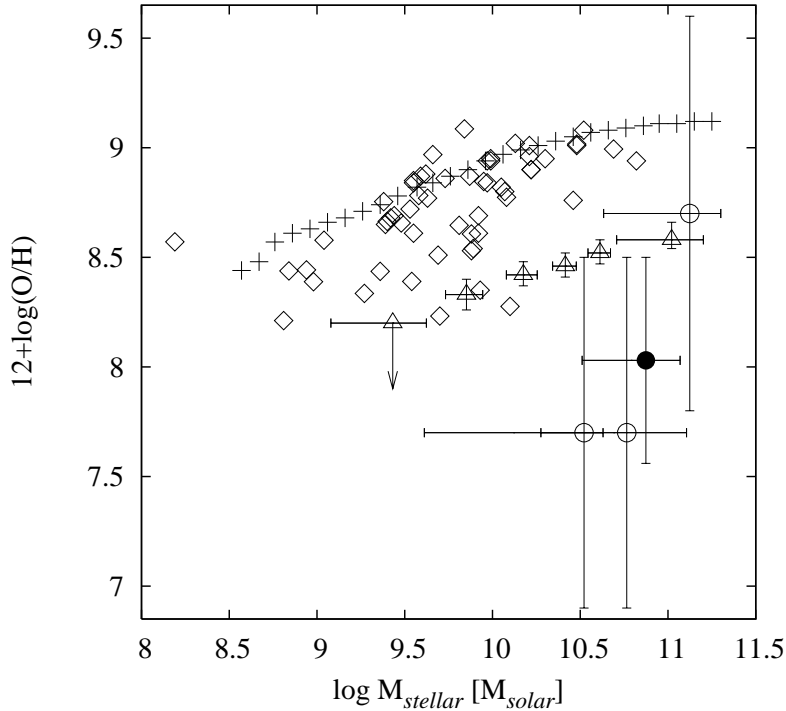}
\end{center}
\caption{Stellar mass-metallicity relation for galaxies at various redshifts. 
Open circles represent our results of 3 luminous LBGs at $z\sim5$, and
filled circle represents the average value of them.
Pluses, diamonds, and triangles show the results for galaxies at
 $z\sim0.1$, $z\sim0.7$ and $z\sim2$, respectively \citep{Tre,Sav,Erb06}.
}
\end{figure}

\bigskip

We are grateful to the FOCAS team, especially Youichi Ohyama, 
and the staffs of Subaru telescope for their supports during our
observation.
We also thank Kiyoto Yabe and Marcin Sawicki
for useful discussions in deriving stellar masses.
We appreciate the anonymous referee for the comments which improved
this paper.
The preparation of the observation was in part carried out on "sb" 
computer system
operated by the Astronomical Data Analysis Center (ADAC) and Subaru
Telescope of the National Astronomical Observatory of Japan.
MA is supported by a Research Fellowship of the Japan Society for the
Promotion of Science for Young Scientists.
KO is supported by a Grant-in-Aid for Scientific Research from
Japan Society for the Promotion of Science (17540216).
This work is supported by the Grant-in-Aid for the 21st Century COE
"Center for Diversity and Universality in Physics" from the MEXT of Japan.

\end{document}